\documentstyle[12pt,epsf]{article}

\renewcommand{\section}[1]{\vspace{6pt} \noindent\mbox{#1} \newline \noindent}
\renewcommand{\subsection}[1]{\vspace{6pt} \noindent\mbox{\underline{#1}} 
\newline \noindent}
\renewcommand{\subsubsection}[1]{\vspace{6pt} \noindent\mbox{\underline{#1}}
\noindent}

\newfont{\sansb}{cmssbx10}
\newfont{\sans}{cmss10}

\begin{document}
{\small OG 9.1.14 \vspace{-24pt}\\}     %The session code
{\center \LARGE SPECTRUM AND CHEMICAL COMPOSITION OF CRs ACCELERATED IN SNRs
\vspace{6pt}\\}
E. G. Berezhko, G. F. Krymsky, V. K. Yelshin and L. T. Ksenofontov
\vspace{6pt}\\ {\it Institute of Cosmophysical Research and Aeronomy, 31
Lenin Ave., 677891 Yakutsk, Russia \vspace{-12pt}\\} 
{\center ABSTRACT\\}
Spectrum and chemical composition of cosmic rays accelerated in supernova
remnants are studied on a basis of kinetic approach. The cosmic ray
transport equation with the Bohm diffusion coefficient has been numerically
solved self-consistently with gas dynamic equations for the underlying flow.
Comparison with observational results gives some indications that galactic
cosmic rays are produced by supernovae shocks which expanded in the relatively low
temperature partially ionized interstellar medium.

\setlength{\parindent}{1cm}
\section{INTRODUCTION}
Supernova remnants (SNRs) are considered as the main source of cosmic rays
(CRs) in the Galaxy. Diffusive shock acceleration process is able to convert
sufficient amount of the explosion energy into CRs and to produce CR
spectrum with necessary shape and amplitude (Berezhko et al., 1996). At the
same time there is no observational confirmation that the nuclear component
of CRs is produced in SNRs. Investigation of CR chemical composition can
give additional evidence whether SNRs are indeed the source of
observed CRs (Berezhko et al.,1995; 1996; Ellison et al., 1997). The most
pronounced aspect of the observed CR chemical composition is the increasing of
relative abundance as function of atomic number $A$. Here we present the
results to study under which conditions the theory fits the observed
features of CR chemical composition.

\section{METHOD}
The description of SNR evolution and CR acceleration is based on the kinetic
approach. It includes (Berezhko et al., 1996) the diffusive transport
equation for the CR distribution function which is solved self-consistently
together with the gas dynamic equations for the underlying flow. The model
includes an assumption about injection of some fraction $\eta \ll 1$ of
superthermal gas particles (ions) at the subshock into the acceleration
regime. By definition gas particles becomes CRs when they reach some minimum
velocity $v_{inj}$ during the thermalization. Therefore the momentum of
injected particles
\begin{equation}
p_{A inj} = A p_{inj}
\end{equation}
is proportional to the atomic number $A$. Here and hereafter variables without
subscript $A$ corresponds to protons.

The expected chemical composition of accelerated CRs is sensitive to the
dependence of the injection rate $\eta_A$ upon the atomic number.
According to the hybrid plasma simulation of quasiparallel collisionless
shock (Trattner and Scholer, 1993) the relative number of particles in the
power-law tail of the spectrum is approximately the same for protons and
$\alpha$-particles. Therefore we have to take
\begin{equation}
\eta_A = \eta .
\end{equation}
We assume that supernovae (SN) shock expands into the uniform interstellar
medium (ISM) with normal chemical composition. \hspace{0.1cm} Therefore
\hspace{0.1cm}the\hspace{0.1cm} relative\hspace{0.1cm} number\hspace{0.1cm}
of \hspace{0.1cm} accelerated \hspace{0.1cm} nuclei\\ $N_{A inj}/N_{inj}=
(\eta_A/\eta)(a_A/a)=a_A/a$ is
determined by the solar system abundance $a_A$ of elements with the atomic
number $A$.

The shape of accelerated CR spectrum essentially depends upon the assumed
CR diffusion coefficient. We use
\begin{equation}
\kappa_A (p_A) = {p_A c^2 \over Q_A e} ,
\end{equation}
which at relativistic energy ($p_A \gg Amc$) coincides with so called Bohm
diffusion coefficient. Here $c$ is the speed of light, $m$ and $e$ are the
mass and charge of protons. The ion charge number $Q_A$ is assumed to be a
function of momentum $p_A$. At the beginning of acceleration, that
corresponds to small momenta $p_A \sim p_{A inj}$, $Q_A$ coincides with the
equilibrium ion charge state in background ISM $Q_{A i}$. At nonrelativistic
energies the acceleration process is extremely fast because of the small
value of the diffusion coefficient. Therefore we assume the constant ion
charge number
\begin{equation}
Q_A(p_A)=Q_{A i} \mbox{\hspace{0.4cm} at \hspace{0.4cm} } p_A \leq Amc.
\end{equation}
The acceleration process becomes progressively less rapid with increasing CR
energy. Therefore at the sufficiently high energies $p_A \gg Amc$ all kind
of ions becomes completely ionized, that means
\begin{equation}
Q_A=Z_A \mbox{ \hspace{0.4cm} at \hspace{0.4cm}}  p_A \gg Amc,
\end{equation}
where $Z_A$ is the nuclear charge number of elements with atomic number $A$.

Due to the rigidity-dependent CR leakage from the Galaxy, the expected CR
spectrum (flux) is connected with the spectrum $J_A(p_A)$ produced in SNR
(source spectrum) by the relation
\begin{equation}
J_{A obs}(p_A) \propto J_A(p_A) \tau(p_A/Z_A),
\end{equation}  
where we assume the power-low rigidity dependent mean CR residence time
\begin{equation}
\tau (R) \propto R^{-\alpha}.
\end{equation}  

\section{RESULTS AND DISCUSSION}
In order to investigate the influence of nonlinear effects produced by the
accelerated CR backreaction and ISM parameters on the CR chemical
composition we have performed calculations for high ($\eta=10^{-3}$) and low
($\eta=10^{-5}$) injection rates and for two typical phases of ISM: warm
(with temperature $T\simeq 10^4$K) and hot ($T\simeq 10^6$K). The initial
ionic charge numbers $Q_{A i}=1$ for all elements in warm ISM and $Q_A$
slowly increases from $Q_{A i}=1$ for $H$ and $Q_{A i}=2$ for $He$ to $Q_{A
i}=9 $ for $Xe$ in the hot ISM (Kaplan and Pikelner, 1979). The value of the
parameter $\alpha$ (see eq.(7)) has been chosen to fit the observed
proton's CR spectrum. 

Calculated hydrogen and helium spectra and CR abundance relative to solar
system abundance versus atomic mass number (enhancement factor)
\begin{equation} e(A,\varepsilon_k) = \left [J_{A
obs}\left({\varepsilon_{A k}\over A}\right) \left /
J_{obs}\left(\varepsilon_k={\varepsilon_{A k}\over
A}\right)\right] \left / \left ({a_A \over a} \right )
\right. \right. . \end{equation}
at kinetic energy per nucleon $\varepsilon_k=\varepsilon_{A k}/A=3$GeV
are compared on Figure 1 with experimental data. The normalization of
calculated spectra was made by fitting the observed hydrogen spectrum at
kinetic energies $\varepsilon_k = 10^2 \div10^4$GeV (the relative
normalization of hydrogen to helium and other nuclei is fixed by the model).
In the case of low temperature ISM our theory is applicable only for
gas-phase ions only for which experimental data are presented on Figure 1b. We
do not consider here the problem of acceleration of refractory elements,
which locked in grains in the low temperature ISM (this interesting question
was considered by Ellison et al, 1997).

One can see from Figure 1 that in the case of low injection, when nonlinear
modification of the SN shock by the CR backreaction is negligible the theory
in contrast to observation predicts much less efficient production of all
kinds of species relative to protons. CR enrichment by heavy elements is not
essential also for the high injection rate ($\eta=10^{-3}$) in the case of hot
ISM. Only in the low temperature ISM, with a low degree of ionization for all
kind of species, the theory predicts essential CR enrichment by heavy
elements if the injection rate is relatively high.

The shape of CR spectrum produced by the modified shock is essentially
different at relativistic and nonrelativistic energies: $J(p) \propto
p^{-\gamma_1}$ at $p \leq mc$ and $J(p) \propto p^{-\gamma_2}$ with
$\gamma_2 < \gamma_1$ at $p\gg mc$ (Berezhko et al., 1996). The higher shock
modification, the larger difference between $\gamma_1$ and $\gamma_2$. Taking
into account the form (3) of the diffusion coefficient, for heavier species we
have $J_A(p_A/A) \propto a_A(p_A/A)^{-\gamma_1}$ at $p_A \leq Q_{A i} mc$
and $J_A(p_A/A) \propto a_A(p_A/A)^{-\gamma_2}$ at $p_A\gg Q_{A i} mc$ that
gives at relativistic energies the enhancement factor
\begin{equation} e(A)  = \left({ A \over Q_{A i}}\right)^{\gamma_1 -
\gamma_2} \left({ A \over Z_A}\right)^{-\alpha} . \end{equation}
The ratio $A/Z_A$ is close to 2 for all elements except hydrogen and
$\alpha$ is about $0.7$ for three considered cases. Therefore for unmodified
shock $\gamma_1 \simeq \gamma_2$ and we have $e<1$ that means lower heavy
element abundance relative to protons. The essential CR
enrichment by heavy species ($e>1$) is expected only under two condition: i)
in the case of low temperature ISM with low ionization state $Q_{A i} \simeq
1 $ and ii) at relatively high injection rate, which provide the efficient
CR production and strong SN shock modification. It is important to note that
according to the theory (Ellison et al., 1997; Trattner and Scholer, 1993;
Malkov and V\"olk, 1995) and experiment in the solar wind (Trattner et al.,
1994) the expected injection rate is not less than $\eta = 10^{-3}$.

\begin{figure}
\epsfxsize=17cm
\epsfbox[47 433 588 737]{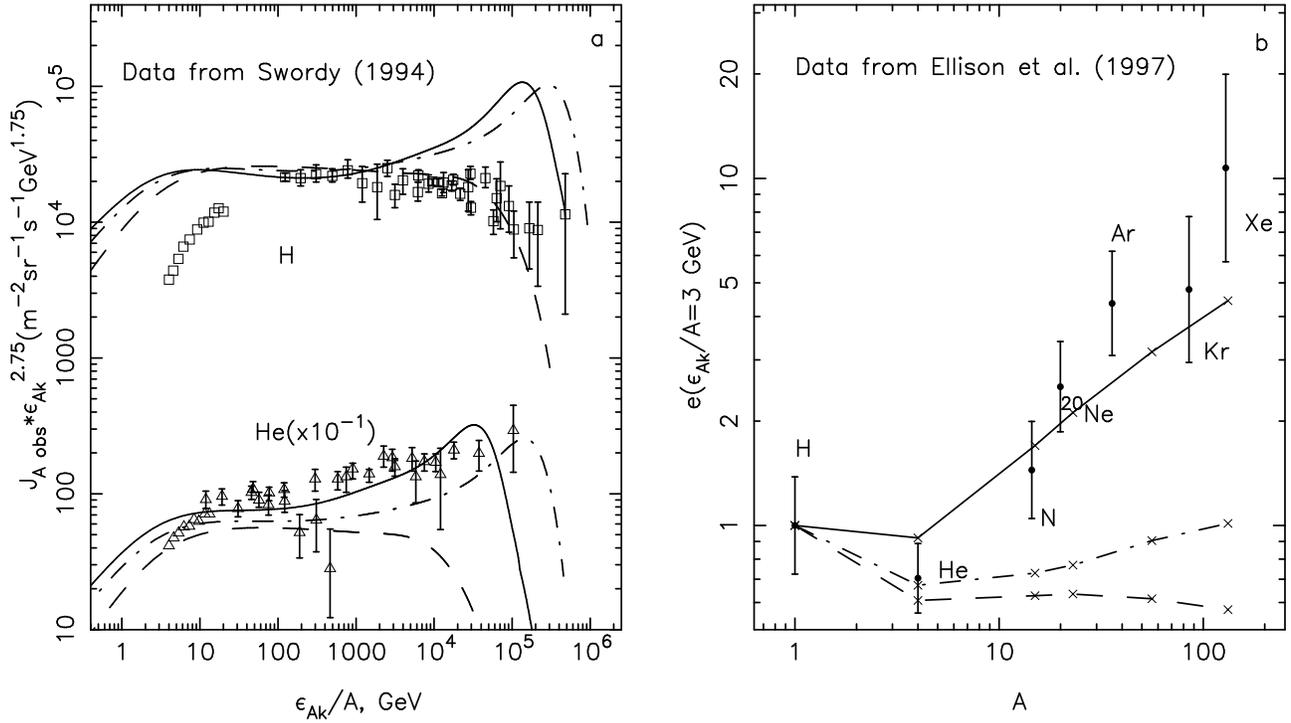} 
%\vspace{19cm}
\caption{CR spectra in kinetic energy per nucleon (a) and CR enhancement
factor versus atomic mass number (b) for ISM temperature $T=10^6$ and
injection rate $\eta=10^{-3}$ (dash-dotted lines), for  $T=10^4$ and $\eta=10^{-5}$
(dashed lines),  for  $T=10^4$ and $\eta=10^{-3}$ (full lines)    }
\end{figure}   

Only under the above two conditions, as one can see from Figure 1b, the theory
satisfactorily agrees with the data. At the same time, as it is seen from
Figure 1a, in the case of high injection the theory predicts a much more harder
hydrogen spectrum at highest energies $\varepsilon_ k > 10^{13}$eV than
observed.

Our CR spectra are also essentially harder compared with plane-wave kinetic
model prediction (Ellison et al. 1997). The main reason of it is that in our
case all phases of SNR evolution contribute to the overall CR spectrum
(Berezhko et al. 1996). The highest energy CRs are produced at the end of
free expansion phase, when the SN shock is especially strong and produces
extremely hard CR spectrum. CRs with intermediate energies
$\varepsilon_k=10^{10} \div 10^{13}$eV are mainly produced during the Sedov
phase, when the SN shock is considerably weaker. As a result the overall CR
spectrum is not of the pure power-law form but has a concave shape. In this
sense the CR spectra presented by Ellison et al. (1997) correspond to the
late Sedov phase, therefore they are essentially steeper than ours which
include contribution of all phases of SNR evolution. 

The possible explanation of the discrepancy between our theory and the
experiment can be connected either with the acceleration or with the
propagation of CRs in the Galaxy. In first case due to some underestimated
physical factors SNR produces a more steeper CR spectrum than is predicted
by our theory. The second possibility is that the actual escape of CRs from
the Galaxy is more complicated than described by simple power-law Eq.(7).

\section{ACKNOWLEDGMENTS}
This work has been supported in part by the Russian Foundation for Basic
Research (97-02-16132).

\section{REFERENCES}
\setlength{\parindent}{-5mm}
\begin{list}{}{\topsep 0pt \partopsep 0pt \itemsep 0pt \leftmargin 5mm
\parsep 0pt \itemindent -5mm}
\vspace{-15pt}
\item Berezhko, E. G., Yelshin, B. K., and Ksenofontov, L. T., {\em ZhETPh},
{\bf 109}, 3 (1996).
\item Berezhko, E. G., Krymsky, G. F., Ksenofontov, L. T., and Yelshin, V. K.,
{\em Proc. 24th ICRC}, {\bf 3}, 392 (1995).
\item Ellison, D. C., Drury, L. O'C. and Meyer, P., Accepted in {\em ApJ}
(1997).
\item Kaplan, S. A. and Pikelner, S. B., in {\em Physics of Interstellar
Medium}, ed. N. G. Bochkarev, pp. 296--356, Nauka, Moscow (1979).
\item Malkov, M. A. and V\"olk, H. J., {\em A\&A}, {\bf 300}, 605 (1995). 
\item Trattner, K. J., M\"obius, E., Scholer, M. et al., {\em JGR}, {\bf
99, A7}, 13389 (1994).
\item Swordy, S., in {\em Proc. 23th ICRC, Invited, Rapporteur and Highlight
Papers }, eds. D. A. Leahy, R. B. Hicks, and D. Venkatesan, p. 243, World
Scientific (1994).
\item Trattner, K. J. and Scholer, M., {\em Ann. Geophysicae}, {\bf 9}, 774
(1993).

\end{list}

\end{document}